# Design of a High-bunch-charge 112-MHz Superconducting RF Photoemission Electron Source

T. Xin[1,2], J. C. Brutus[1], Sergey A. Belomestnykh[#,&,1,2], I. Ben-Zvi[1,2], C. H. Boulware[3], T. L. Grimm[3], T. Hayes[1], Vladimir N. Litvinenko[1,2], K. Mernick[1], G. Narayan[1], P. Orfin[1], I. Pinayev[1], T. Rao[1], F. Severino[1], J. Skaritka[1], K. Smith[1], R. Than[1], J. Tuozzolo[1], E. Wang[1], B. Xiao[1], H. Xie[4], A. Zaltsman[1]

[1] Brookhaven National Laboratory, Upton, NY 11973-5000, U.S.A.
[2] Stony Brook University, Stony Brook, NY 11794, U.S.A.
[3] Niowave, Inc., Lansing, MI 48906, U.S.A.
[4] Peking University, Beijing, China

*Abstract*

High-bunch-charge photoemission electron-sources operating in a continuous wave (CW) mode are required for many advanced applications of particle accelerators, such as electron coolers for hadron beams, electron-ion colliders, and free-electron lasers (FELs). Superconducting RF (SRF) has several advantages over other electron-gun technologies in CW mode as it offers higher acceleration rate and potentially can generate higher bunch charges and average beam currents. A 112 MHz SRF electron photoinjector (gun) was developed at Brookhaven National Laboratory (BNL) to produce high-brightness and high-bunch-charge bunches for the Coherent electron Cooling Proof-of-Principle (CeC PoP) experiment. The gun utilizes a quarter-wave resonator (QWR) geometry for assuring beam dynamics, and uses high quantum efficiency (QE) multi-alkali photocathodes for generating electrons.

## INTRODUCTION

With SRF becoming the technology of choice for many new particle accelerators, its functionality is being extended from accelerating structures to other systems [1]. One of these new uses of SRF is in high-brightness CW photoinjectors [2,3]. SRF gun technology, while still in the development stage, promises to enable generating electron beams with high-bunch-charge, high-brightness, and high-average-current. The concept of a quarter-wave resonator (QWR) SRF gun was proposed at BNL for the electron cooling of hadron beams in RHIC [4,5]. QWRs can be made sufficiently compact even at low RF frequencies/long wavelengths. The long wavelength allows us to produce long electron bunches, thus minimizing the effects of space charge. QWR guns are especially well suited for generating high-brightness and high-bunch-charge beams as they can operate with high electric field at the cathode with an optimum emission-phase close to 90°, i.e. on the crest; see [6] for more details.

To achieve high luminosity, future electron-ion colliders (EICs) will require advanced techniques of electron cooling to reduce the emittance of ion beams. BNL is building an experiment to demonstrate the feasibility of coherent electron cooling [7] for its version of the EIC, called eRHIC [8]. The CeC PoP experiment [9] will need high-charge electron bunches, from 1 to 5 nC per bunch, with a repetition rate of 78 kHz, to cool one of the ion bunches in the relativistic heavy-ion collider, RHIC. A 112-MHz superconducting quarter-wave resonator electron-gun cryomodule was designed and built in a collaboration between BNL and Niowave, Inc. as part of testing the concept of coherent electron cooling [10]. The cryomodule, with its associated cathode preparation and insertion and other sub-systems will serve as the CeC PoP injector. The gun is designed to deliver electrons with a kinetic energy of up to 2 MeV. Electrons are generated by illuminating a high QE $K_2CsSb$ photoemission layer with a green laser operating at a wavelength of 532 nm. Fig. 1 shows the layout of the 112 MHz gun, recently installed in the RHIC tunnel and commissioned with a beam, generating 3 nC bunches at 1.7 MeV. The installation and commissioning results, along with a comprehensive analysis of the gun's performance are reported elsewhere [6,11]. This article is devoted to a detailed description of the design of the SRF photoemission source.

---

[#] Corresponding author, sbelomes@fnal.gov
[&] Present address: Fermilab, PO Box 500 MS 316, Batavia, IL 60510-5011, USA

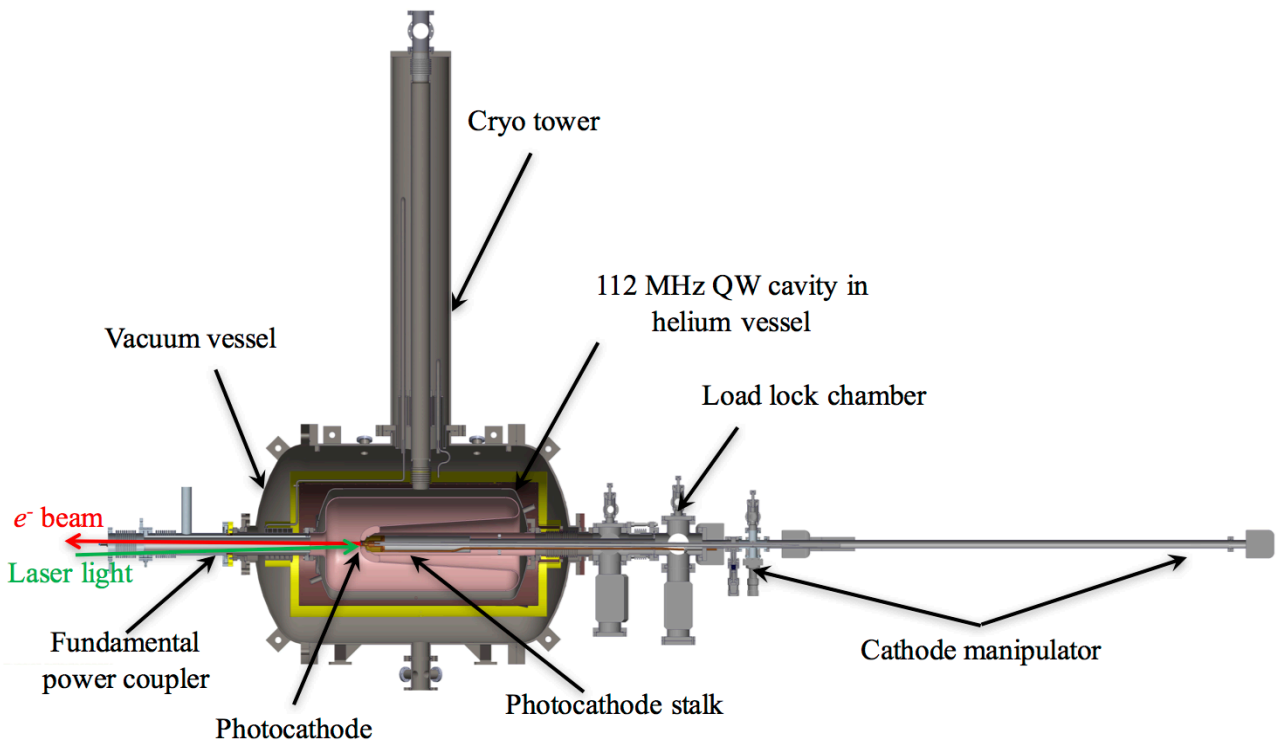

Figure 1: Cross-section of the 112 MHz SRF gun (elevation view). The laser light is entering the gun through a hollow fundamental power coupler, which also provides and exit path for the electron beam.

## CAVITY DESIGN AND FABRICATION

The cavity's geometry was designed and optimized using Superfish [12], a 2D code for simulating RF structures. Fig. 2 presents the Superfish geometry used to calculate the parameters of the RF cavity; they are summarized in Table 1, alongside other parameters of the gun. Since the fields near the cathode will appear nearly electrostatic, a Pierce-type geometry (recessed cathode, as shown in the figure) can be used to focus the beam and partially compensate for the space charge of the high-charge bunches. The Pierce electrode shape positions higher electric fields away from the cathode, but these field levels were controlled during the design process. The cathode's axial position can be adjusted, offering an opportunity to explore its effect on the beam's quality.

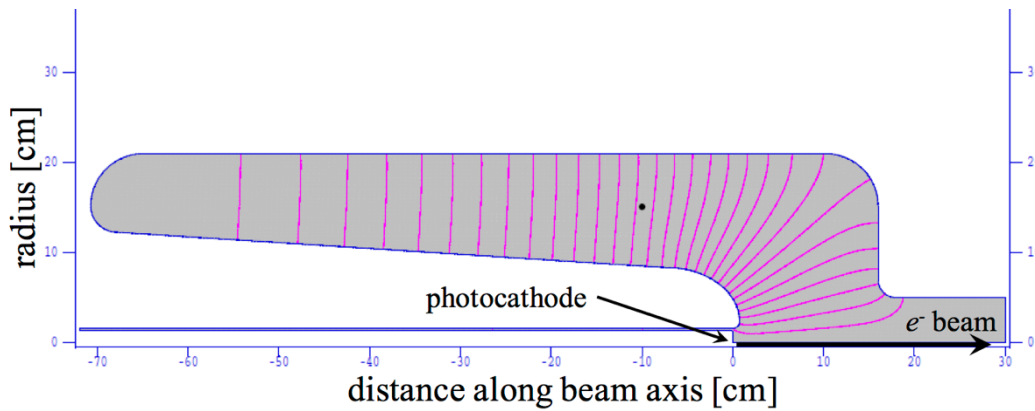

Figure 2: 112 MHz cavity geometry used in the Superfish calculations.

Table 1: Parameters of the 112 MHz SRF gun

| Parameter | Value |
|---|---|
| Frequency | 112 MHz |
| *R/Q* (linac definition) | 126 Ohm |
| Geometry factor, *G* | 38.2 Ohm |
| Quality factor $Q_0$, w/o cathode insert | $> 3.5 \times 10^9$ |
| Operating temperature | 4.5 K |
| Accelerating voltage $V_{acc}$ | 1.5 to 2.0 MV |
| $E_{pk}/V_{acc}$ | 19.1 m$^{-1}$ |
| $E_{pk}/E_{cath}$ | 2.63 |
| $B_{pk}/V_{acc}$ | 36.4 mT/MV |
| Bunch charge | 1 to 5 nC |
| Normalized emittance | < 5 mm·mrad |
| Bunch repetition frequency | 78 kHz |
| Available RF power | 2 kW |
| Cavity length | 1.1 m |
| Cavity diameter | 0.42 m |
| Accelerating gap | 0.1525 m |
| Beam pipe's aperture | 0.1 m |

The cavity was fabricated from high RRR (residual resistivity ratio) niobium by electron-beam-welding pre-formed parts together. Welding was performed in four cycles: Welding the niobium beam tubes; welding the inner and the outer conductors from pre-formed halves; welding the sub-assembly; completing all niobium welding of the cavity. Fig. 3 shows the complete cavity. The beam tubes are terminated by stainless-steel CF flanges. A stainless-steel helium vessel was welded on to the cavity before chemically etching the cavity's inner surface; it also acted as a cooling jacket. Preparation of the cavity followed a "standard" recipe for preparing the surface of SRF cavities and included buffered chemical polishing (BCP 1:1:2) and high-pressure rinsing with particulate-free de-ionized water, as described, for example, in [13].

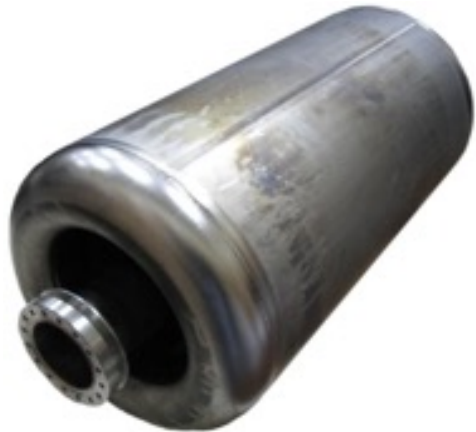

Figure 3: Electron-beam welded 112 MHz niobium cavity.

Fig. 4 depicts the layout of the 112 MHz QWR inside the gun's cryomodule. The cavity in its helium vessel is surrounded by a blanket of superinsulation (not shown), a thermal shield cooled by gaseous helium at 12 to 16 K, another blanket of superinsulation, and finally a magnetic shield. The cryomodule's vacuum vessel is made of stainless steel and meets the requirements of the ASME Pressure Vessel Code. It comprises the main cylindrical part and two end caps, bolted together. In addition to the cavity, the layout in Fig. 4 shows a fundamental RF-power coupler (FPC) and a cathode stalk, both of which will be described in detail in subsequent sections. The cathode stalk is designed to minimize the transverse field on the cathode surface, as well as loss of RF power on the stalk itself. The coaxial-type fundamental power coupler is designed to couple the RF power into cavity through the beam pipe exiting the caavity. Its center conductor (coupling tube) is hollow, thus allowing beam to go through it. By adjusting the penetration of the coupling tube, one can tune the cavity's resonant frequency as well as adjust the coupling strength. The beam pipes inside the cryomodule connecting the niobium cavity with the vacuum vessel are made of stainless steel with a copper-plated inner surface to reduce RF losses. There are heat intercepts connected to the heat shield in the middle of both pipes. The pipe enclosing the cathode stalk incorporates bellows to accommodate thermal contraction during cooldown.

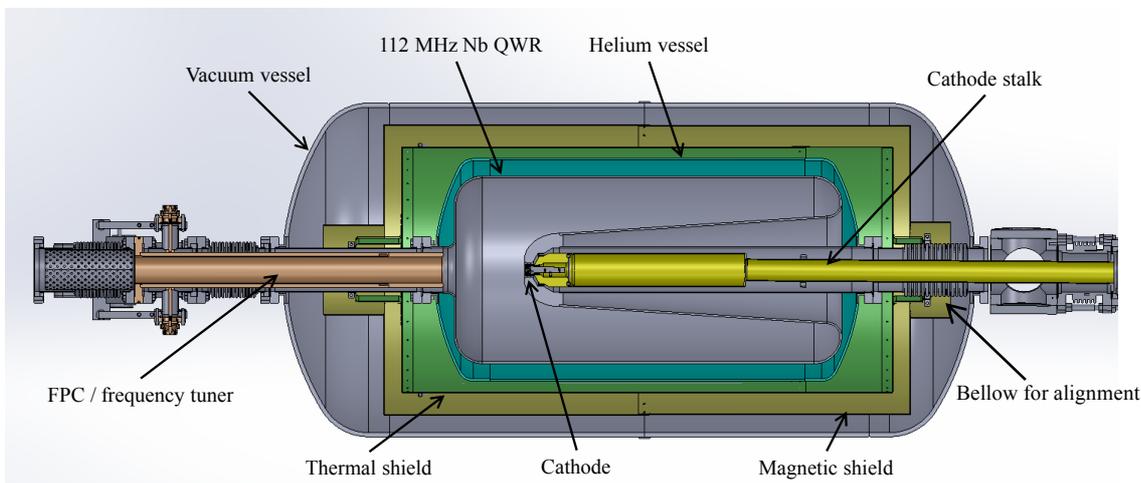

Figure 4: Plane view layout of the 112 MHz QWR with the FPC and cathode stalk inside the gun's cryomodule.

## CATHODE DESIGN, PREPARATION, AND INSERTION SYSTEM

The photocathode QE drops significantly with temperature, as reported in [14,15] and the literature cited therein. At low temperature, the cathode work function increases due to the bandgap enlargement. As a result, the QE significantly decreases at a given wavelength of 532 nm. To avoid this problem, the 112 MHz gun was designed with a cathode operating at room temperature. The $K_2CsSb$ photoemission layer is deposited on the front surface of a small, 20-mm diameter, molybdenum puck (cathode in Fig. 4). This cathode resides inside a cathode stalk and is inserted with a specially designed cathode-manipulation system. The cathode stalk is permanently installed inside the gun. It serves two functions: 1) It thermally insulates the cathode from the cold center conductor of the QWR; and 2) It provides an RF choke joint between the cathode and the cavity. The latter minimizes the transverse electric field near the cathode, thus improving the beam's dynamics. Detailed design considerations regarding the cathode stalk, the cathode's preparation and manipulation are discussed below.

### Design of the cathode stalk

The cathode stalk's design principle is similar to that described in [16]. The stalk is a hollow normal-conducting structure made of copper-plated (25 μm) stainless steel operating at room temperature. It does not have direct physical contact with the cold center conductor of the cavity, thus reducing the leakage of heat to the cavity only allowing the exchange of radiative heat. To reduce its emissivity, the stalk is coated with a very thin (about 1 µm) layer of gold. A Rexolite® "spider" is utilized to center the cantilevered stalk inside the cavity. The spider is indicated in Fig. 6 as "center support".

However, this design allows the RF field to propagate out of the cavity into a gap between the stalk and the center conductor. A half-wavelength stalk shorted at the far end makes a choke filter, thus reducing the penetration of the RF field, and minimizing the voltage drop between the cathode and the cavity's center conductor. The gap between the stalk and nose cone at the entrance of the cavity is only 3.56 mm. The stalk's length was shortened to account for the capacitance created by this small gap. The resulting electric field distribution near the cathode is shown in Fig. 5. By integrating along the black solid line, we can obtain the voltage drop between the cathode and cavity, which is only 110 V at an accelerating voltage of 2 MV.

If the cathode stalk were straight, i.e. without an impedance mismatch step in the middle, the resulting RF power losses there would be 36 W. Further reduction of RF losses on the stalk, from 36 W to 20 W, is achieved by introducing an impedance mismatch in the middle of the stalk (as shown in Figs. 4 and 6), which creates a quarter wavelength impedance transformer. The remaining heat is removed by cooling water. Another advantage of the impedance mismatch is the reduction of power loss on the alignment bellows, which could overheat if the RF loss stayed as high as the original number. Fig. 6 illustrates the stalk assembly. Table 2 compares RF losses on different parts of the stalk assembly between a uniform stalk design and a stalk with an impedance transformation step.

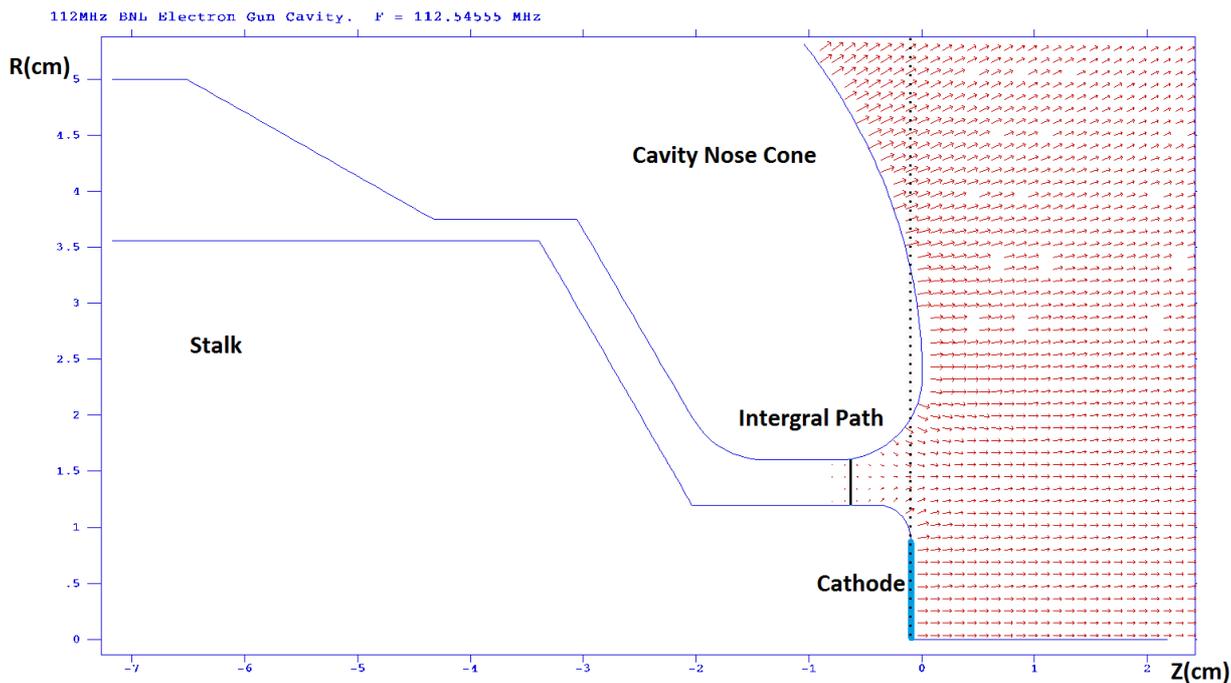

Figure 5: Distribution of the electric field near the cathode's surface and the nose cone (the gap between the cathode and stalk is not shown.)

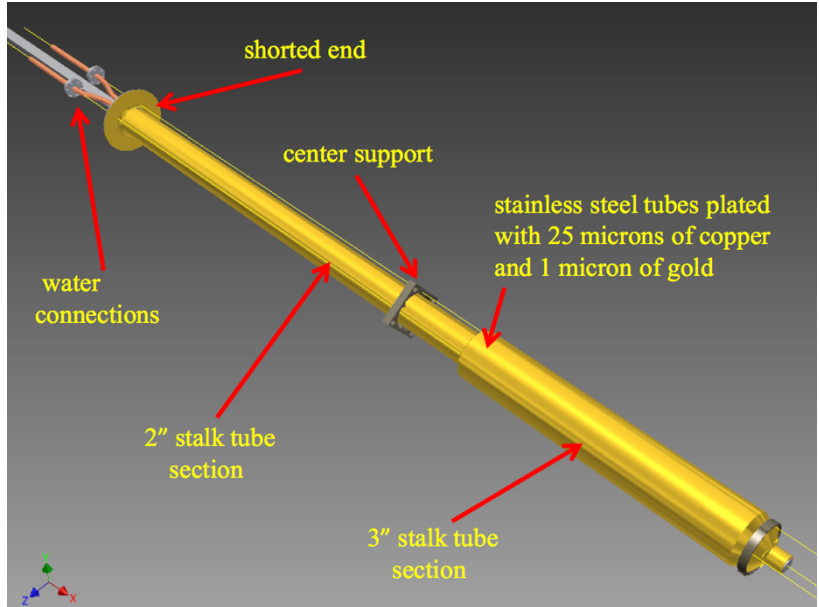

Figure 6: Cathode stalk assembly.

Table 2: Comparison of RF losses on a uniform stalk and a stalk with impedance-transforming step at an accelerating voltage of 2 MV

| Power losses | Uniform stalk | Stalk with impedance-transformation step |
| --- | --- | --- |
| on the stalk (Au) | 36 W | 20 W |
| on the bellows (Cu) | 28 W | 7 W |
| on the Nb pipe | 120 µW | 31 µW |

*Cathode preparation*

A high QE cathode is needed for generating high-charge bunches. For the required high repetition frequency, it is difficult to obtain high laser pulse energy. Thus having a high QE cathode makes it practical to generate bunch charges in the nC range [17]. We chose $K_2CsSb$ because of its high QE, ~10% at 532 nm [17], and long lifetime, ~30 hours in a $10^{-10}$ torr-scale vacuum environment while delivering high average beam current [18]. In addition, this wavelength makes it easier to shape laser pulses spatially and temporally. The cathodes are deposited at a deposition chamber described in [19] and shown in Fig. 7. The deposition chamber is equipped with a residual gas analyzer (RGA), a quartz-crystal monitor, and multiple viewports for observation and laser irradiation of the cathode. Alkali-metal dispenser sources (6 mg potassium, and 10.8 mg cesium from SAES getter) and antimony (99.999% purity pellets from Goodfellow) are used for the deposition. These sources are inserted through gate valves into the main chamber during evaporation, but are isolated from each other and from the main chamber to prevent cross-contamination of the sources. There is a boron-nitride resistive heater mounted beneath the cathode holder to control the substrate temperature from 20°C to 500°C.

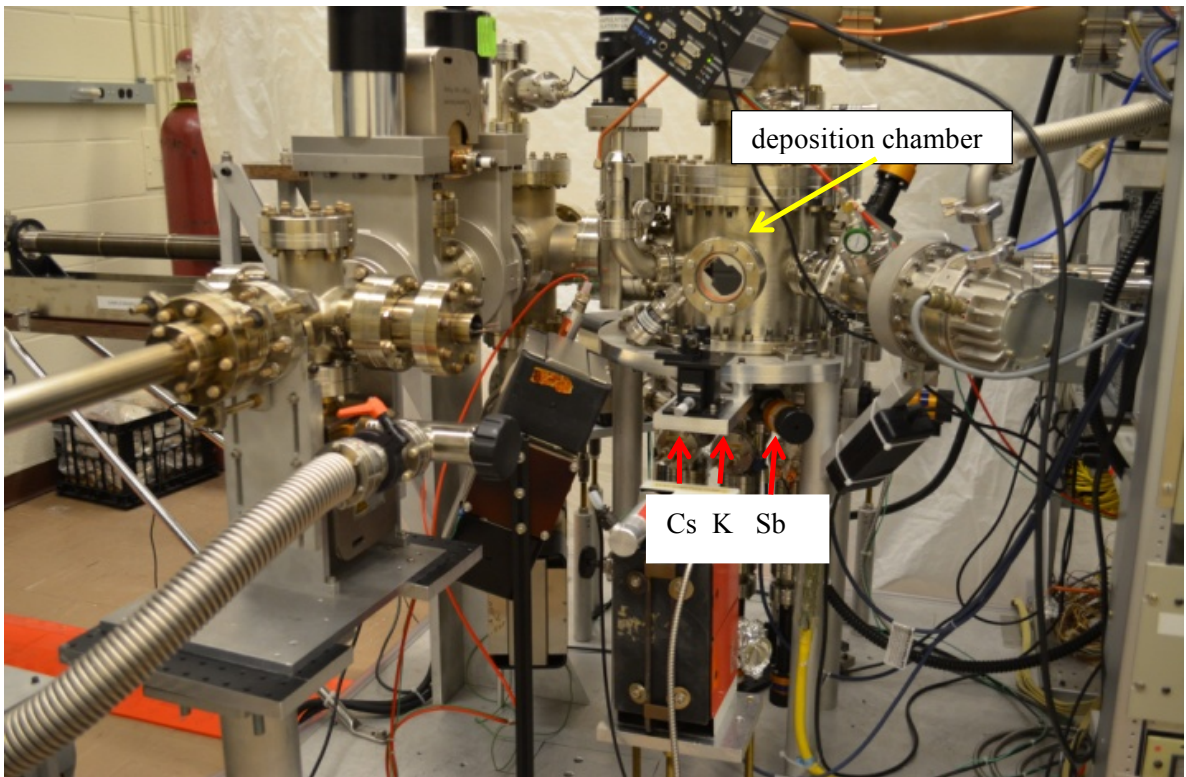

Figure 7: Multi-alkali deposition system for the 112 MHz gun.

The cathode material is deposited on a 20-mm diameter pucks made of high purity molybdenum. The puck's front surface is diamond-polished to 2 nm rms roughness, to reduce the roughness-induced emittance [20,21]. Fig. 8 shows the molybdenum pucks before depositing the cathode. We use a standard cathode- deposition procedure, which is discussed in [22]. Two cathodes were prepared. one with an initial QE at 6.5%, and the other at 8%. After four weeks' storage inside a portable in-vacuum injection storage chamber called garage, and transferring it into a load-lock space, one cathode had QE of 4% before inserting it into the gun [15]. The garage vacuum, normally at $1\cdot10^{-10}$ torr, increases up to $5\cdot10^{-10}$ torr during operation of the manipulator.

The entire cathode substrate was exposed to alkali vapor in the multi-alkali deposition chamber. The multi-alkali coated surface has low work function and as a result – high secondary emission yield. This created favorable conditions for multipacting in a narrow gap between the side surface of the puck and the stalk even at very low, 110 V or less, voltages existing in this gap. Multipacting would not be possible there for a bare molybdenum surface. Because of the low voltage in the gap, the impact energy of electrons is also low, which makes it very difficult to condition the surface. Thus, we used excimer laser treatment to remove the active cathode material from the puck's edge. This method was studied in 2013 for the 704 MHz SRF photocathode deposition system [16]. After cleaning the cathode by laser ablation, its QE significantly degraded, to less than 0.001%, due to outgassing during the laser ablation process. Typical pressure during cleaning process was in the $10^{-8}$ to $10^{-7}$ torr range. As a multi-alkali cathode cannot sustain high QE in a vacuum worse than $10^{-9}$ torr due to surface contamination by adsorbed gases, the cathode surface required "rejuvenation". We warmed up the cathode to 80°C for two hours to "rejuvenate" it. A QE of 0.8% was measured after the cathode stabilized in a preparation chamber. This QE was confirmed during the 112 MHz gun beam test [6].

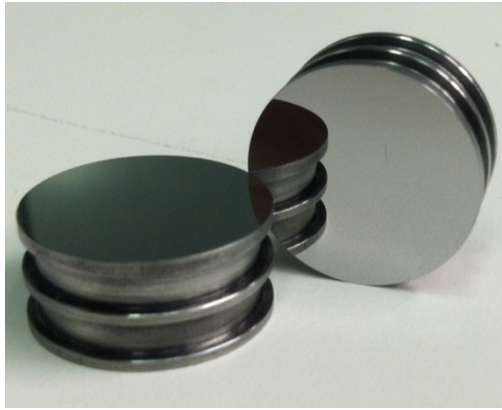

Figure 8: Photograph of polished molybdenum pucks before deposition of the photoemission layer.

*Cathode insertion system*

The cathode insertion system is depicted in Fig. 9. The garage, which is the portable in-vacuum injection system in Fig. 9, with up to three cathodes inside is connected to the gun via a load lock. The cathode's diagnostic system allows us to measure the cathode's QE before inserting it inside the gun using a two-meter-long magnetic manipulator. The cathode-end assembly attached to the manipulator arm is shown in Fig. 10. The manipulator arm has three centering standoffs with rolling ceramic wheels, which prevent damage to the cathode on insertion, and limit the generation of particulates. Two grooves on the cathode puck allow manipulation with special spring "forks". Finally, gold-plated RF spring finger contacts connect the puck with the inner surface of the stalk ensuring that electromagnetic field does not propagate inside the cathode stalk.

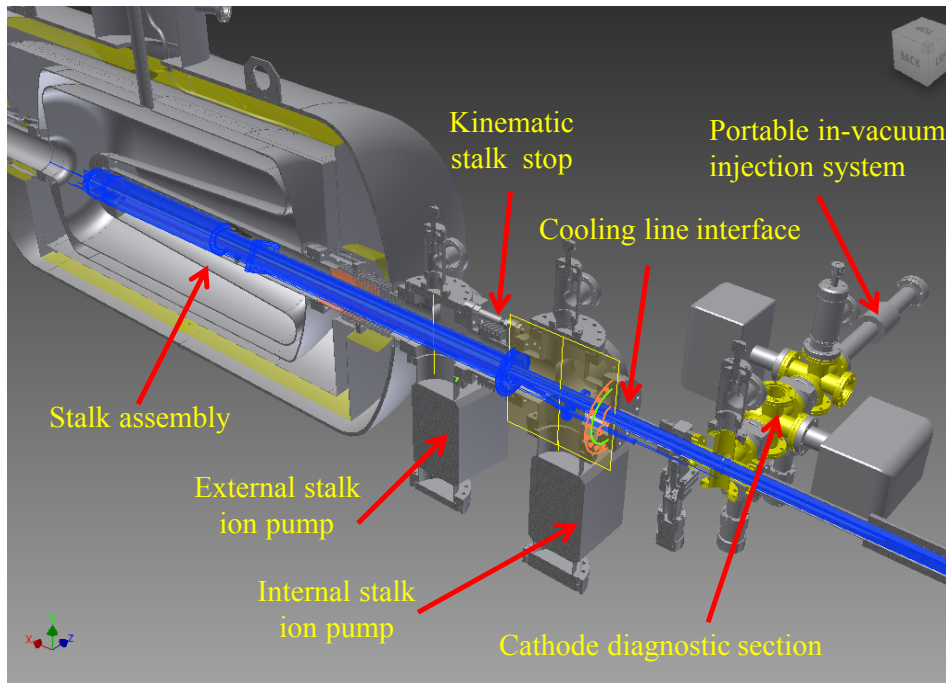

Figure 9: Cross-sectional view of the cathode-insertion system.

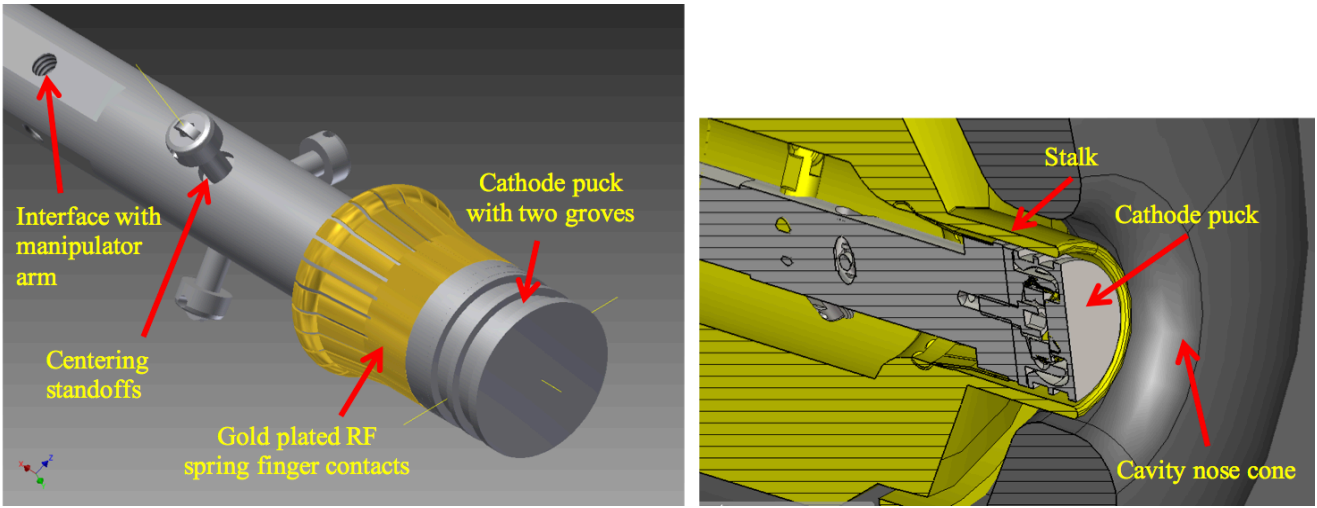

Figure 10: Cathode end assembly: The cathode puck attached to the end of the manipulator arm (left); Sectional view of the puck inside the stalk (right).

## FUNDAMENTAL POWER COUPLER

The fundamental power coupler is a coaxial-type coupler similar to one used in the NPS gun [16]. The coupler is attached to beam exit port of the SRF gun. Its hollow center conductor, or coupling tube, allows the passage of the beam. By adjusting the penetration of the coupling tube, we can tune the cavity's resonant frequency as well as adjust the coupling strength. Fig. 11 illustrates the FPC attached to the SRF gun port. Table 3 lists the main parameters of the fundamental power coupler.

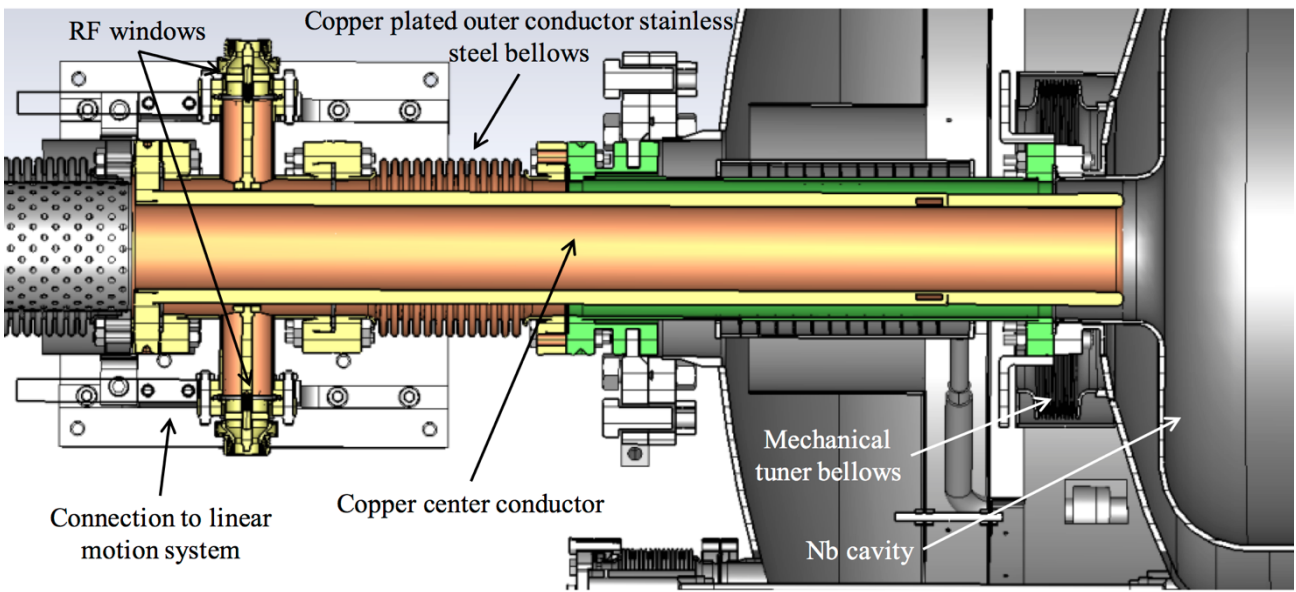

Figure 11: Cross-section of the FPC attached to the beam-exit port.

Table 3: Parameters of the fundamental power coupler

| | |
|---|---|
| Frequency tuning range with FPC | 4.5 kHz |
| Travel range | 40 mm |
| $Q_{ext}$, min. | $1.25 \cdot 10^7$ |
| Max. RF power loss on the inner conductor | 899 W |
| Max. RF power loss on the outer conductor | 555 W |

*RF design of the FPC*

To allow it to change the cavity's frequency, the FPC [23] is designed as a resonant quarter-wavelength structure slightly detuned down, by ~2.8 MHz, from the cavity's resonance. There are significant RF fields inside the structure, generating much larger losses that those inside the cavity. Fig. 12 shows the variation of the gun's FPC external quality factor with the position of the coupling tube relative to the cavity (blue line), calculated using CST Microwave Studio [24]. The range is from $5 \cdot 10^6$ to $1.2 \cdot 10^8$.

Also shown in Fig. 12 are plots of the external quality factor under critical coupling condition for three values of the bunch charge, assuming that there is no parasitic detuning due to microphonic noise. The optimal external $Q$ is calculated using the following equation [25]:

$$Q_{ext\_opt} = \frac{Q_{cav}}{1+\frac{P_{beam}}{P_{cav}}} = \frac{Q_{cav}}{1+\frac{f_b \cdot q_b \cdot V_{acc} \cdot \cos\varphi_b}{V_{acc}^2/(R/Q \cdot Q_{cav})}}, \qquad (1)$$

where $Q_{cav}$ is the cavity quality factor, which accounts for losses in the cavity walls, the cathode stalk, and the FPC; $P_{beam}$ is the power delivered to the beam; $P_{cav}$ is the power dissipated in the cavity, cathode stalk, and FPC; $f_b$ is the bunche repetition rate; $q_b$ is the bunche charge; and $\varphi_b$ is the bunch phase. The bunch is assumed to be on-crest for the plots shown in Fig. 12. Points where the blue curve intersects other curves indicate the coupling tube position at which critical coupling condition is achieved. As is evident from Fig. 12, depending on the FPC's position and bunch charge we would need $Q_{ext}$ from $10^7$ to $10^8$ for optimal matching.

For the CeC experiment, the SRF gun frequency has to be a harmonic of the 40 GeV/u ion-beam revolution frequency in RHIC, which is approximately 78 kHz. To cover this range, the gun frequency will be roughly tuned to the desired frequency by a mechanical tuner while the FPC is near the critical coupling position. The tuner mechanism is attached to the cryostat and the adjustment is manual. The fine-tuning is provided by adjusting position of the FPC, which tuning range is shown in Fig. 13. We also installed a phase shifter in the external circuit between two feedthroughs. This system, which is currently not in use, will allow for fine adjustment on the coupling when FPC is used to tune the gun frequency.

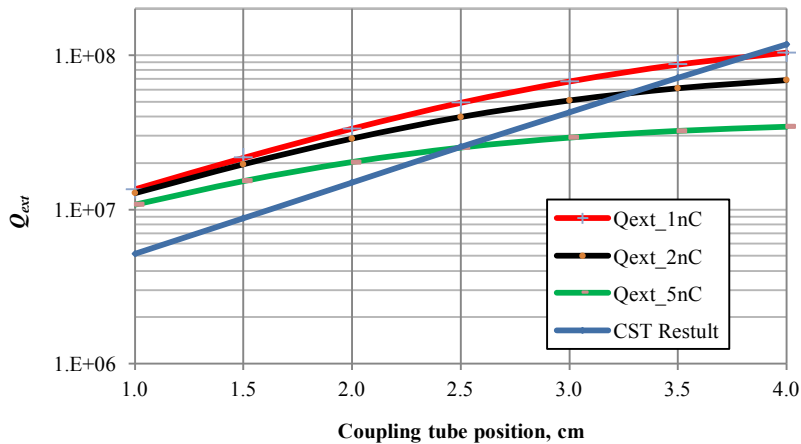

Figure 12: $Q_{ext}$ versus position of the coupling tube.

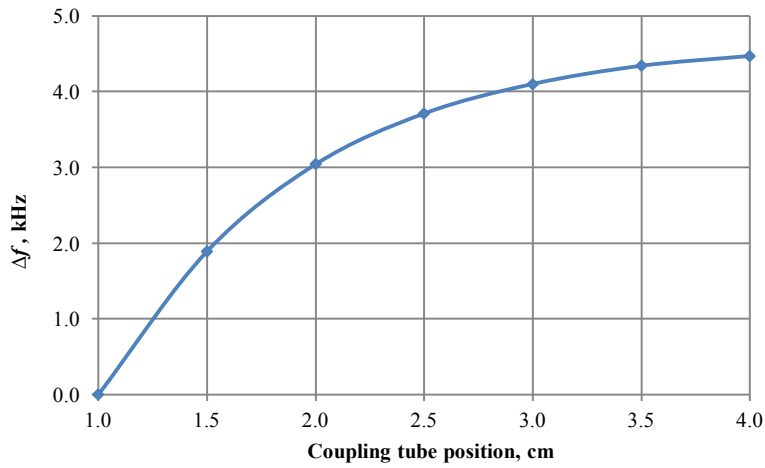
Figure 13: Frequency tuning by FPC.

*FPC mechanical design and fabrication*

A detailed layout of the FPC assembly is shown in Fig. 14. The FPC comprises a center (or inner) conductor welded to a flange, two bellows sections, and an antenna section with three small ports for attaching ultra-high vacuum (UHV) RF windows (feedthroughs) and vacuum-system components. All components are made of stainless steel with copper-plated surfaces facing the RF field to reduce losses. The inner conductor consists of two sections welded together. A shorter section near the cavity does not have channels for water cooling, while the longer section has those channels. The assembly is supported at the antenna section by a linear motion system with a travel range of 40 mm. The FPC is powered symmetrically via two UHV 7/16 DIN coaxial RF feedthroughs, as shown in Fig. 15.

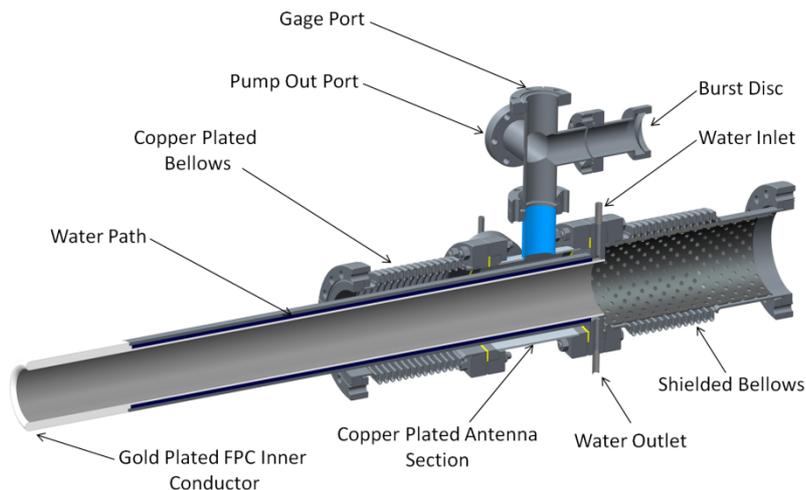
Figure 14: Detailed layout of the 112 MHz FPC.

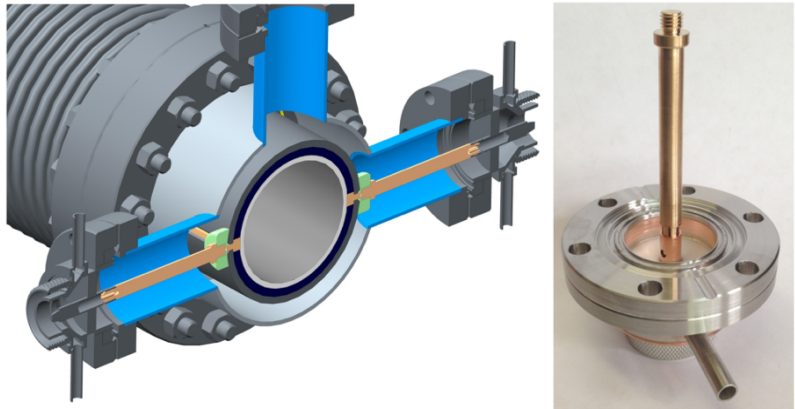

Figure 15: Sectional view of two RF feedthroughs connected to the FPC (left); 7/16 DIN coaxial UHV feedthrough (right).

One challenge with the FPC design was to provide adequate cooling for removing heat due to RF power-losses on the surfaces of the inner- and outer-conductors when the gun is operating. The outer conductor includes the copper-plated bellows. Structural and thermal analyses, using ANSYS [26], confirmed the FPC's structural stability and also were used to calculate the deflection and maximum temperature. The distribution of RF power dissipated on the FPC's inner- and outer-conductors is shown in Table 5.

Table 4: Distribution of RF power loses on the FPC

| RF Power Loses | Value |
| --- | --- |
| On inner conductor section with water cooling | 894.3 W |
| On inner conductor section without water cooing | 4.3 W |
| On bellows of outer conductor | 554.6 W |
| On flange | 21.9 W |

To find out if cooling was needed for the bellows on the outer conductor of the FPC, ANSYS was used to determine the maximum temperature on the overall system. Heat fluxes were applied to the inner surfaces according to the values in Table 4. A heat-transfer coefficient of 15 W/m$^2$K corresponding to a natural convection by air was applied on the outer surfaces. A maximum temperature of 460 K was observed at the depth of the convolution. The result from ANSYS for the natural convection cooling of the outer conductor of the FPC is shown in Fig. 16. As a focusing solenoid is mounted on top of the bellows, thus restricting natural convection, a forced-air-flow system is installed in this area to improve cooling.

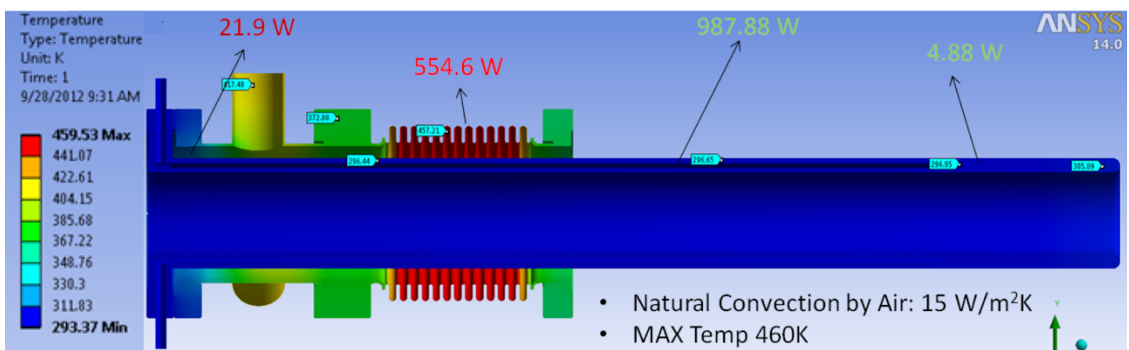

Figure 16: Temperature profile on the FPC's outer surface.

The inner conductor of the FPC is cooled using de-ionized low-conductivity water. Fig. 14 shows that the inner conductor has a shorter and a longer section. The shorter section is designed to stiffen the structure and space out the water cooling channels from the 4 K region of the QWR. The longer section consists of two water channels, one in the upper half, and the other in the lower half. Both channels are connected towards the cavity end of the inner conductor. The water flows through the upper half first and exits through the lower half of the conductor with a calculated temperature difference of approximately 1 K between the inlet and outlet. Table 5 gives the cooling system parameters. The ANSYS results indicate that the FPC's inner conductor will reach a maximum temperature of 305 K (Fig. 17), and a maximum deflection of 0.001 inch (25.4 µm).

Table 5: Cooling system parameters

| Parameters | Value |
|---|---|
| Water volume flow rate, $Q$ | 4 gpm (2.252 l/s) |
| Pressure drop | < 2 psi (13.8 kPa) |
| Local heat transfer coefficient, $h$ | 3217.4 W/m$^2$K |
| Inlet-outlet temperature difference | 0.88 K |

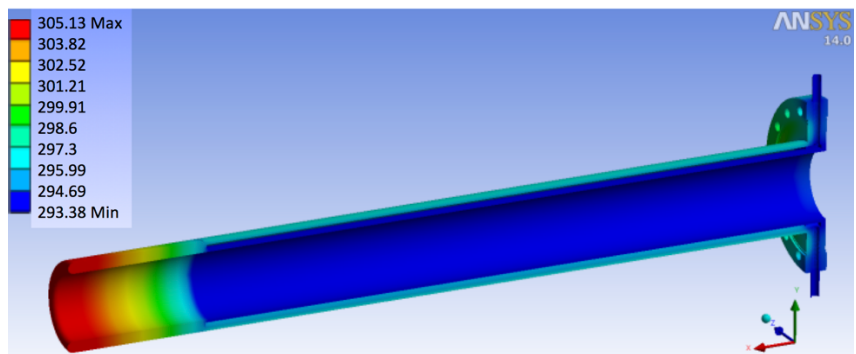

Figure 17: Temperature profile on the FPC inner conductor.

The copper-plated and the shielded bellows shown in Fig. 14 were designed to provide 40-mm travel with a low spring constant of 84 lbs/in to reduce the moments on the linear motion system. The plated bellows, shown in Fig. 18, have a minimum of a 25 µm thick layer of copper on the inside of the convolutions.

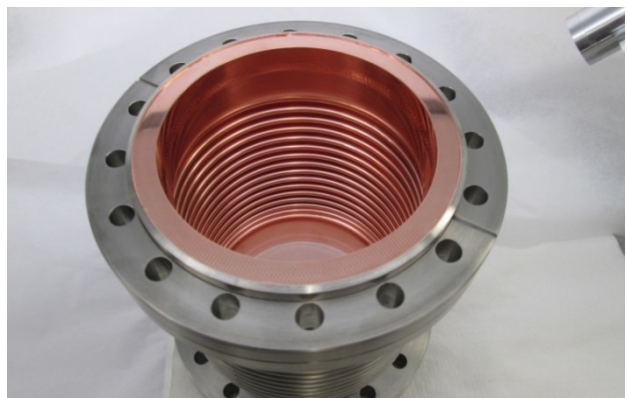

Figure 18: Copper-plated bellows.

The inner conductor of the FPC was fabricated from 304 stainless-steel. The final assembly, shown in Fig. 19, was copper plated with a 25 μm thick layer, then plated with a 1 μm thick layer of gold to maintain a mirror-like surface finish and bring the emissivity to as low as 0.02. This was done to reduce heat radiation to the 4 K environment of the 112 MHz superconducting QWR gun, and to prevent the cooling water from freezing in case of a failure.

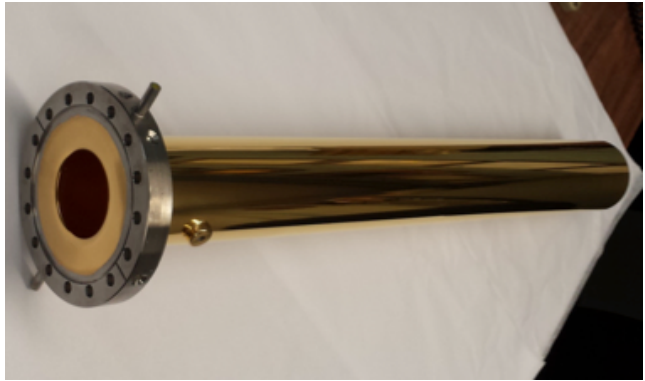

Figure 19: 112 MHz FPC gold-plated inner conductor.

The linear stage motion system for the FPC shown in Fig. 20 was designed and fabricated in a collaboration between BNL and Ibex Engineering. The design uses a granite base to reduce deflection and vibration. The linear stage was successfully tested and met all the requirements listed in Table 6.

Table 6: Parameters of 112 MHz FPC linear stage

| Parameters | Value |
| --- | --- |
| Travel range | 40 mm |
| Response frequency | 10 Hz |
| Position resolution | 500 nm |
| Positioning accuracy | ± 5 μm |
| Repeatability (no backlash- and hysteresis-free) | < ±500 nm |
| Straightness over travel range | ± 0.5 μm |
| Flatness over travel range | ± 0.5 μm |
| Speed | > 1 mm/s |
| Acceleration | > 60 mm/s$^2$ |
| Position readback resolution | < 1 μm |
| Moment M1 transverse to the direction of travel | 250 lbs·in (28.2 N·m) |
| Moment M2 in the direction of travel | 1000 lbs·in (113 N·m) |

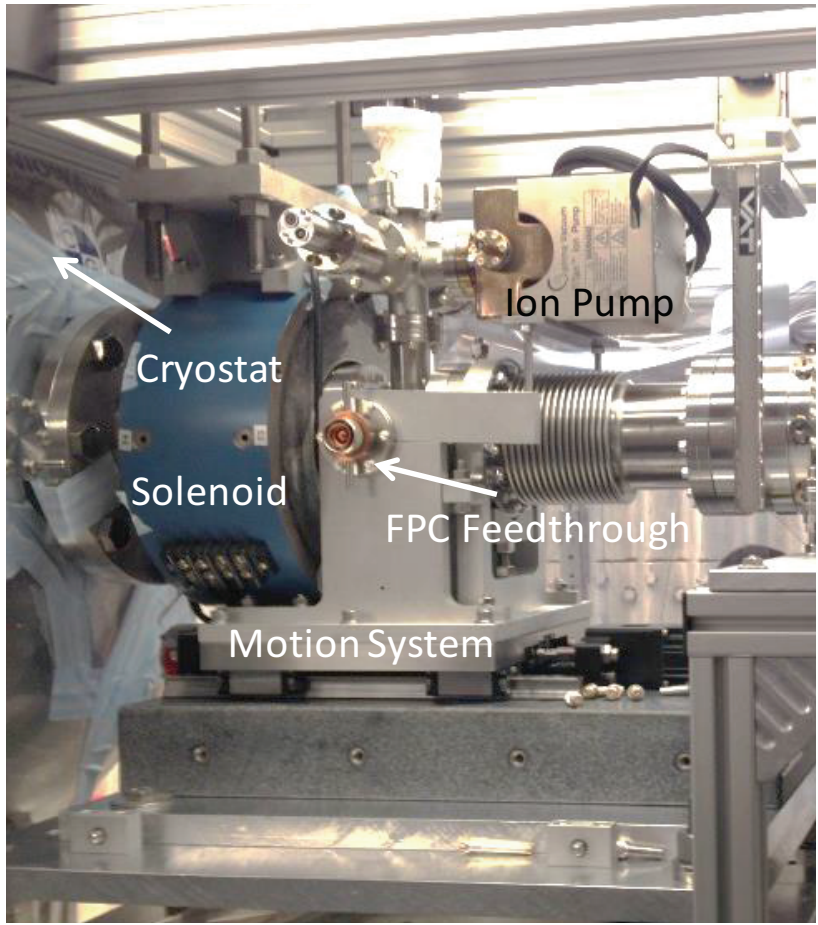

Figure 20: Photo of the FPC installed on the 112 MHz QWR SRF gun with its linear motion system. The focusing solenoid (blue) covers the copper-plated bellows. A small ion pump is installed on top of the antenna section.

To protect the FPC and the cathode system, an automated water interlock system was successfully tested and installed. The interlock system monitors pressure, water flow, and temperature of both the FPC and the cathode system. It is designed to protect the FPC and cathode system by preventing RF power to the cavity without the minimum water flow requirement. Since the FPC and the cathode system are near proximity of the 4 K region, the system also is designed to automatically blow out the water line using helium gas to prevent the water from freezing inside the cavity in case of failure of water flow.

**MULTIPACTING SIMULATIONS**

Multipacting (MP) analysis of the cavity, the FPC, and cathode stalk was done using Track3P [27] and our own GPU-based particle tracking code. Both simulations found several MP barriers, as shown in Fig. 21, where the enhanced counter function indicates the total number of secondary electrons generated by a single initial electron after a given number of impacts, which is 100 impacts in our case. The first barrier, located inside the cavity, appears when the gun voltage gets into the range of 40 to 50 kV. The second barrier emerges at approximately 200 kV gun voltage, and continues to exist until about 650 kV. It is located inside the FPC. Finally, the third barrier is inside the cathode stalk and the corresponding gun voltage is from 600 kV to 1 MV. The MP locations are shown in Fig. 22. All MP barriers were observed during the gun's commissioning.

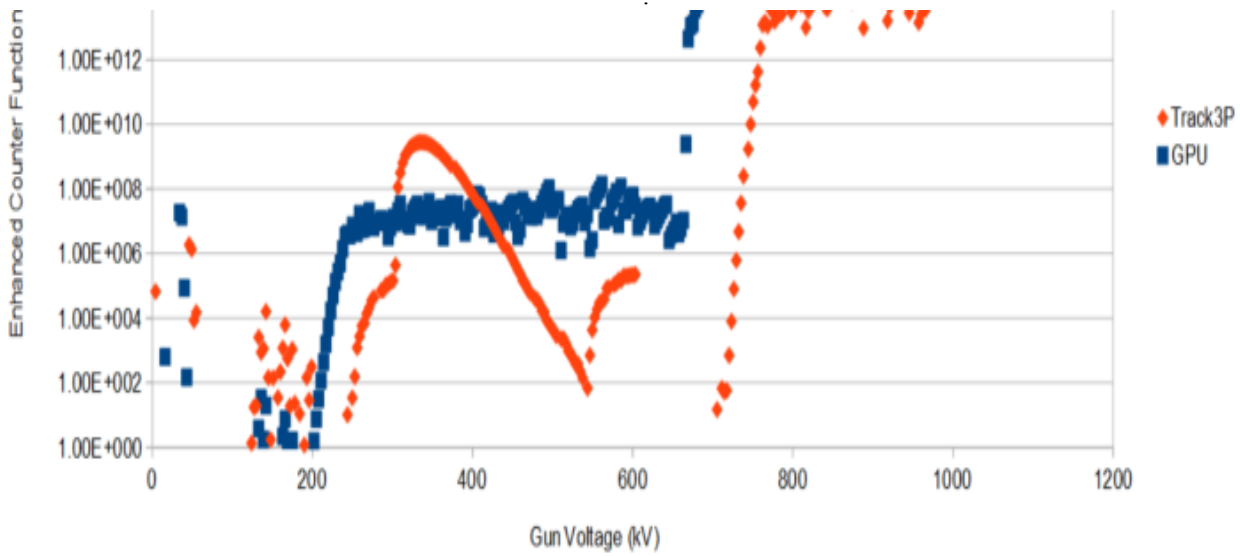

Figure 21: Enhanced counter function given by GPU code and Track3P.

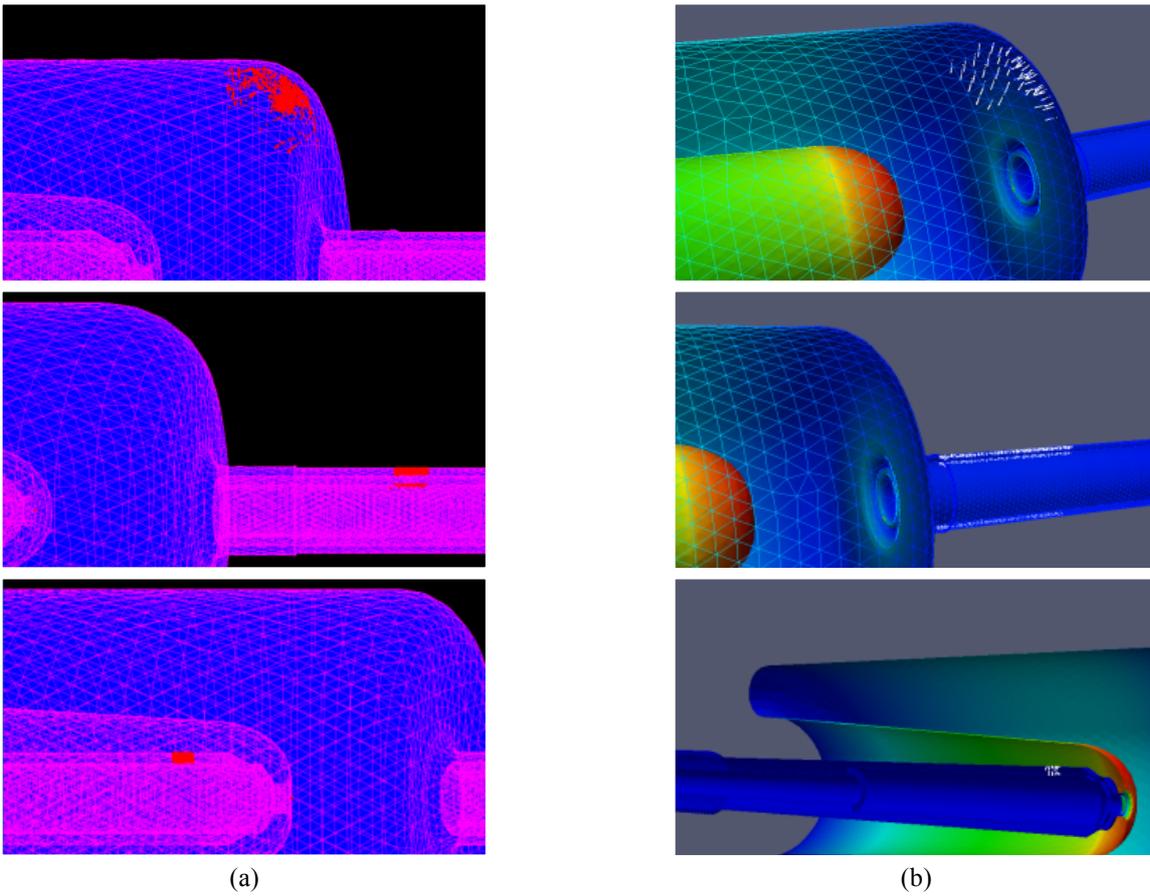

Figure 22: Location of MP in the 112 MHz injector given by (a) GPU code, resonant particles shown as red dots, and (b), Track3P resonant particles shown as white dots. The pairs of pictures from top to bottom show the MP at gun voltage equal to 40 kV, 200 kV, and 600 kV respectively.

## COMMISSIONING RESULTS

Our SRF gun is part of the RHIC accelerator complex. It uses the RHIC cryogenic system for operation. This synchronizes its operation with RHIC runs. The gun commissioning began in the fall of 2014 (RHIC Run 15) without a photocathode puck. Several multipacting zones, consistent with computer simulations, were encountered. They were cleaned out after several days of conditioning, and never presented a problem afterwards. The gun eventually reached the voltage level where its performance was limited by field emission (FE). High-power pulsed processing allowed us to proceed further and the gun has reached stable operation in CW mode at 1.3 MV, limited by a very high cryogenic load due to FE. A cavity voltage as high as 1.8 MV could be reached in the pulsed mode. Afterwards, helium conditioning [25] was applied to the cavity, allowing us to reduce the FE dark current, and improve the cavity's performance to 1.7 MV in CW, and 2 MV in a pulsed mode. The first beam test of the SRF gun commenced with a temporary laser, which has a repetition rate limited to 5 kHz. The measured quantum efficiency (QE) of the photocathode of ~0.8% was in good agreement with expectations. The first measured bunch charge was around 1.35 nC due to the space-charge limitation. By increasing the laser spot size at the cathode to approximately 2.5 mm, we increased the maximum extracted bunch charge to just above 3 nC. Operating at 5 kHz repetition rate, we generated an average beam current of more than 15 µA. More details on the installation and commissioning results along with comprehensive analysis of the gun performance can be found elsewhere [6,11]. After the upgrading the laser system, we expect no problems generating 78 kHz bunches at the same energy and bunch charge as with the temporary laser.

At the time of the submission, we have restarted operation of the gun for RHIC Run 16. After successful re-conditioning, the first beam from the photocathode was generated. A normalized emittance of 0.5 mm·mrad (preliminary) was measured at 0.5 nC and 1.2 MV. Detailed studies of the gun performance during this RHIC run will be published elsewhere.

## SUMMARY

We have designed a high-charge superconducting RF photoemission electron source, based on a 112 MHz quarter-wave resonator. The gun employs high QE multi-alkali photocathodes deposited on small molybdenum pucks. The cathodes are prepared in a deposition chamber and transported between the chamber and the gun in a "garage" under ultra-high vacuum. A half-wavelength cathode stalk allows the cathode to operate at room temperature. The gun's fundamental RF power coupler is adjustable, and serves also as a fine-frequency tuner. The SRF gun was fabricated and installed in the RHIC tunnel as part of the CeC PoP experiment. During first round of commissioning, we managed to overcome multipacting and field emission to a level allowing us to demonstrate first beam with a record for such devices bunch charge of 3 nC.

## ACKNOWLEDGEMENT

The authors acknowledge financial support from the DOE Office of Nuclear Physics, Facilities and Project Management Division, "Research and Development for Next Generation Nuclear Physics Accelerator Facilities Program" FOA (DE-FOA-0000632) and from National Science Foundation (award PHY-1415252).